\documentclass[a4paper,12pt]{article}
\usepackage{graphicx,amssymb,bm,latexsym}
\newcommand{\bea}{\begin{eqnarray}}
\newcommand{\ena}{\end{eqnarray}}
\newcommand{\vs}[1]{\vspace{#1 mm}}

\renewcommand{\a}{\alpha}
\renewcommand{\b}{\beta}
\renewcommand{\c}{\gamma}

\newcommand{\e}{\epsilon}
\newcommand{\s}{\sigma}
\renewcommand{\t}{\theta}
\newcommand{\dsl}{\pa \kern-0.5em /}

\newcommand{\shalf}{\frac{1}{2}}
\newcommand{\pa}{\partial}

\newcommand{\nn}{\nonumber\\}
\newcommand{\p}[1]{(\ref{#1})}

\begin{document}

\begin{titlepage}

\begin{flushright}
hep-th/0603215
\end{flushright}

\vs{10}
\begin{center}
{\Large\bf Matrix String Description of Cosmic Singularities
in a Class of Time-dependent Solutions}
\vs{15}

{\large
Takayuki Ishino\footnote{E-mail address: ishino@het.phys.sci.osaka-u.ac.jp}
and
Nobuyoshi Ohta\footnote{E-mail address:
ohta@het.phys.sci.osaka-u.ac.jp.
Address after 31 March 2006:
Department of Physics, Kinki University,
Higashi-Osaka, Osaka 577-8502, Japan}}\\
\vs{10}
{\em Department of Physics, Osaka University,
Toyonaka, Osaka 560-0043, Japan}

\vs{15}
{\bf Abstract}
\end{center}
\vs{5}

A large class of time-dependent solutions with $\shalf$ supersymmetry
were found previously. These solutions involve cosmic singularities at early
time. In this paper, we study if matrix string description of the singularities
in these solutions with backgrounds is possible and present several examples
where the solutions can be described well in the perturbative picture.

\end{titlepage}
\newpage
\renewcommand{\thefootnote}{\arabic{footnote}}
\setcounter{page}{2}

\section{Introduction}

Study of time-dependent solutions in string theories is an important
subject for its application to cosmology and understanding our
spacetime~\cite{time1}-\cite{null}. Quite often, these solutions
involve spacetime singularities. At the same time the theory becomes
strong coupling and we cannot make definite statement there
because perturbative picture breaks down. It is a major challenge
to understand if these singularities are resolved or not, and if so how.

The solutions with partial supersymmetry (BPS solutions) may be
helpful in this respect because these allow us to discuss nonperturbative
regions of our spacetime. It is known that the requirement of unbroken
partial supersymmetry restricts the solutions to
those with null or time-like Killing spinors~\cite{GP}.

Recently such time-dependent solutions have been studied in the linear
dilaton background in the null direction with $\frac12$
supersymmetry and various extensions have been considered~\cite{bb1}-\cite{bb14}.
It has been argued that it is possible to map the theory to the dual
matrix theory and then to matrix string theory in type IIB picture~\cite{matrix}
where we can use S-duality to put the theory to the weak coupling region.
In this way, it has been suggested that the problem of singularity
in the spacetime is under control in this setting, at least for
the linear dilaton backgrounds~\cite{bb1}.

The solution studied in \cite{bb1} does not involve any background in the forms
which are present in the string theories. However it has not been clear
if the results are restricted to only this very special solution or they
are valid for other related solutions with possible backgrounds.
It is thus important to study how large class of solutions allow such
description of singularities in terms of matrix strings.
In a previous paper~\cite{IKO}, we have found a large class of solutions
of this type in M-theory. However we have not discussed if
this class of solutions have matrix string description though it is
expected that this is the case. The purpose of this paper is to fill
the gap and present a more detailed discussion of this problem.

This paper is organized as follows.
In sect.~2, we first summarize our solutions and propose matrix string
description.
In sect.~3.1, starting from the discrete light-cone quantization,
we show how our solutions can be mapped to the matrix string theory
in the weak coupling region, and  in sect.~3.2 discuss the range of
validity of this approach. In sect.~4, we present several interesting
examples. Sec.~5 is devoted to summary and discussions.

\section{Solutions and the corresponding matrix string theories}

Let us start with a short review of our solutions in 11D supergravity~\cite{IKO}.
The metric of our background is
\bea
ds_{11}^2=-2 e^{2u_0} du dv + \sum_{i=1}^{9} e^{2u_i} (dx^i)^2,
\label{met}
\ena
where all metrics are functions of $u$ and $v$.
We also have the four-form background
\bea
F = (\pa_u E du + \pa_v E dv)\wedge dx^1 \wedge dx^2 \wedge dx^3.
\label{back}
\ena
Upon dimensional reduction to 10 dimensions, we have
\bea
ds_{11}^2=e^{-\frac{2}{3}\phi}ds_{10}^2+e^{\frac{4}{3}\phi}(dx^{9})^2, \nn
ds_{10}^2 = -2 e^{2v_0} du dv + \sum_{i=1}^{8} e^{2v_i} (dx^i)^2,
\label{10D}
\ena
where $\phi$ is a dilaton given by $\phi=\frac{3}{2} u_9$
(and $u_i=v_i-\frac13 \phi,~i=0,1,\cdots,8$).

Within this ansatz, we found that the requirement of remaining supersymmetry
restricts the solutions to only $u$-dependent functions, and
the only condition that must be satisfied is
\bea
\sum_{i=1}^{8} v_i'' + \sum_{i=1}^{8} (v_i')^2 -2 \phi''
+ \frac{1}{2} e^{-2(v_1+v_2+v_3-\phi)}(E')^2 =2 v_0' (\sum_{i=1}^{8} v_i'-2\phi'),
\label{bpse}
\ena
where prime denotes a derivative with respect to $u$, and
the number of remaining supersymmetry is $\frac12$.

Equation~\p{bpse} is an ordinary
differential equation for 11 functions $v_i \; (i=0,\cdots,8), \phi$ and $E$.
We can regard eq.~\p{bpse} as determining $E$ for given metrics.
This in fact gives a very large class of solutions in $D=11$
supergravity, generalizing those discussed in Ref.~\cite{bb1}.

In the simplest solution with the zero four-form, which is the linear dilaton
in the null direction, the string coupling is given by
\bea
g_s = e^{-Q u}
\ena
for positive $Q$. In this string picture, there appears a singularity
in the infinite past $u \to -\infty$, but we see that the string
coupling diverges there. So we cannot say anything definite in this picture.
The suggestion is that with 16 supercharges, it is possible to map the
solution to matrix theory which gives us nonperturbative definition
of the theory~\cite{bb1}. However, this is not enough since the theory
is still in the strong coupling region where we do not know how to
understand the behaviors of the theory. We can then make T-duality
transformation to map the theory to matrix string in type IIB
setting~\cite{matrix} where we can further use S-duality to put the theory
in the weak coupling region. In this way the singularity region gets
a well-defined description. Our question is whether this is true for
our more general class of solutions.

The corresponding matrix string theory is expected to be described by
a $(1+1)$-dimensional super Yang-Mills theory with 16
supercharges~\cite{matrix}.
The action contains eight matrix-valued fields $X^i$ corresponding to
the transverse bosonic coordinates and eight spinor coordinates $\t^\a$:
\bea
S &=&  \frac{1}{2\pi l_s^2}\int d \tau {d\s }\;
\mbox{Tr}\Big(-\frac{1}{2}(D_{\mu} X^i)^2 + \t^T \c_{\mu}D^{\mu} \t
-g_s^{2}l_s^4\pi^2F_{\mu\nu}^2 \nn
&& \qquad +\; \frac{1}{4\pi^2g_s^2l_s^4}[X^i,X^j]^2
+ \frac{1}{2\pi g_s l_s^2}\t^T\c_i[X^i,\t]\Big),
\label{sms}
\ena
where $\mu=\tau,\s$ are the flat world-sheet indices and the coupling constant
is given by
\bea
g_{YM}=\frac{1}{g_s l_s},
\ena
and the contraction rules for the indices of the matrices are determined
by the metrics~\p{10D}. In sect.~\ref{derivation} and examples in sect.~4,
we will give the precise definition.
In the light-cone gauge, the world-sheet time coordinate $\tau$ can be
chosen to be proportional to the spacetime null coordinate $u$ or $X^+$
since we can choose $v_0(u)=0$ using reparametrization in $u$.
However we will keep $v_0(u)$ in the following discussions.

\section{Matrix string description}

\subsection{Derivation}
\label{derivation}

In the discrete light-cone quantization, we make the light-like identification
\bea
v \sim v+R,
\ena
and focus on a sector with light-cone momentum $p^+ = \frac{2\pi N}{R}$.
We define the theory with this identification as a limit of a space-like
compactification
\bea
(v, X^1) \sim (v+R, X^1+\e R),
\label{id}
\ena
where we will eventually take the limit $\e \to 0$.
We are interested in the Lorentz transformation which further puts the
identification to
\bea
(X^-,X^1)\sim(X^-,X^1+\e R).
\label{ident}
\ena
The ``Lorentz transformation''
\bea
u &=& \e X^+, \nn
v &=& \frac{X^-}{\e} + \frac{X^1}{\e}
+ \frac{1}{2\e}\int_0^{X^+} dX^+ e^{2v_0(\e X^+)-2v_1(\e X^+)},  \nn
x^1 &=& X^1 + \int_0^{X^+} dX^+ e^{2v_0(\e X^+)-2v_1(\e X^+)}, \nn
x^i &=& X^i, \qquad i=2,\cdots,8,
\label{tf2}
\ena
puts the background in the form
\bea
ds_{10}^2 &=& -2 e^{2U_0(X^+)} dX^+ dX^- + \sum_{i=1}^{8} e^{2U_i(X^+)} (dX^i)^2,\nn
\phi &=& \phi(\e X^+),
\label{intmet}
\ena
where
\bea
U_i(X^+) \equiv v_i(u) = v_i(\e X^+), \quad
i=0, \cdots, 8,
\label{uis}
\ena
and with the identification~\p{ident}.

We note that the radius of compactification in $X^1$ is small,
but becomes large if we make T-duality transformation.
After T-duality in the $X^1$ direction and then S-duality, we get
\bea
ds_{10}^2 &=& e^{\Phi} \Big( -2 e^{2U_0(X^+)} dX^+ dX^-
+ e^{-2U_1(X^+)} (dX^1)^2 + \sum_{i=2}^{8} e^{2U_i(X^+)} (dX^i)^2\Big), \nn
\Phi &=& -\phi(\e X^+)+U_1(X^+) + \log{r}, \quad
\Big(r\equiv \frac{\e R}{2\pi l_s} \Big),
\label{st2}
\ena
with the identification
\bea
X^1 & \sim & X^1 +  \frac{l_s}{r}.
\ena
This is now a theory of D1-branes in a background where the string
coupling is weak near the big-bang.

One can repeat the analysis of the ground state of the D1-brane theory,
and fluctuations around this ground state~\cite{bb1}. Using the bosonic part of the
Dirac-Born-Infeld action
\bea
S_{D1} = -\frac{1}{2\pi l_s^2}\int d\tau d\sigma e^{-\Phi}
\sqrt{-\mbox{det}(\pa_{\a}X^{\mu} \pa_{\b}X^{\nu}G_{\mu \nu}+2\pi l_s^2F_{\a \b})},
\ena
with the metric~\p{st2}, we examine classical solutions.
Set $X^2,\cdots,X^8,F_{\tau \sigma}$ to zero, and write
\bea
X^1=\frac{l_s}{\e R} \sigma,
\label{cs0}
\ena
so that we have the periodicity $\s \sim \s +2\pi l_s$.
Assuming that $X^\pm$ depend only on $\tau$, the action $S_{D1}$ reduces to
\bea
S_{D1} &=& -\frac{1}{2\pi l_s^2}\int d\tau d\sigma \Big[\frac{l_s}{\e R}
e^{U_0(X^+)-U_1(X^+)}\sqrt{2\pa_{\tau}X^{+} \pa_{\tau}X^{-}} \Big].
\ena
The nontrivial field equations following from this are
\bea
0 &=& \sqrt{2\pa_{\tau}X^{+} \pa_{\tau}X^{-}} (U_0'(X^+)-U_1'(X^+))
e^{U_0(X^+)-U_1(X^+)}\nn
&& -\pa_{\tau}\Biggl(e^{U_0(X^+)-U_1(X^+)}
\sqrt{\frac{\pa_{\tau}X^-}{2\pa_{\tau}X^+}}\Biggr),
\label{X^+} \nn
0 &=& \pa_{\tau}\Biggl(e^{U_0(X^+)-U_1(X^+)}
\sqrt{\frac{\pa_{\tau}X^+}{2\pa_{\tau}X^-}}\Biggr).
\label{X^-}
\ena
These can be satisfied if we take
\bea
e^{2U_0(X^+)-2U_1(X^+)}\pa_{\tau}X^+=\pa_{\tau}X^-.
\ena
We then consider the classical solution obtained by solving
\bea
e^{2U_0(X^+)}\pa_{\tau}X^{+}=\frac{l_s}{\sqrt2 \e R},~~~
e^{2U_1(X^+)}\pa_{\tau}X^{-}=\frac{l_s}{\sqrt2 \e R}.
\label{cs}
\ena
These determine the relation between $X^\pm$ and $\tau$.

We choose the gauge in which $X^+$ and $X^1$ are fixed to the above classical
solution in~\p{cs0} and \p{cs}, and consider the fluctuation
\bea
X^-=X_0^-+\sqrt2\; y,
\ena
where $X_0^-$ is the classical value given by \p{cs}.
We find the fluctuation to the second order
\bea
S_{D1} &=& -\frac{1}{2\pi l_s^2}\int d\tau d\sigma \Big[\Big(\frac{l_s}{\e R}\Big)^2
e^{-2U_1(X^+)}+\Big(\frac{l_s}{\e R}\Big)\pa_{\tau}y
+\frac{e^{2U_1(X^+)}}{2}\Big((\pa_{\sigma}y)^2-(\pa_{\tau}y)^2\Big)\nn
&& +\sum_{i=2}^8 \frac{e^{2U_i(X^+)}}{2}\Big((\pa_{\sigma}X^i)^2
-(\pa_{\tau}X^i)^2\Big)-8\pi^4 l_s^4e^{2\phi(\e X^+)}F_{\tau \sigma}^2+\cdots\Big].
\label{last}
\ena
The second term can be dropped since it is a total derivative.
Comparing this action and the matrix string \p{sms}, we find
that the bosonic coordinates $X^i,i=1,\cdots,8$ in \p{sms}
precisely agree with $y$ and $X^i,i=2,\cdots,8$ in \p{last} if we use the
classical solutions given by~\p{cs} for the background.
The coupling constant on $F_{\mu\nu}$ is also written in terms of
the dilaton $\phi(u)=\phi(\e X^+)$. Thus we conclude that the theory can be
mapped to matrix string.

\subsection{Regime of validity}

We repeat the analysis of the range of validity of the matrix
string description~\cite{bb1}.
Consider a scalar field $T$ with mass $m$.
The equation of motion in our background~\p{10D} is
\bea
0 = \Big(2e^{-2v_0} \pa_u \pa_v -\sum_{i=1}^8 e^{-2v_i}\pa_{x^i} \pa_{x^i}
-2(\phi '+v_0')e^{-2v_0}\pa_v +m^2\Big)T,
\label{T}
\ena
where prime denotes derivative with respect to $u$.
A basis of solutions is
\bea
T(u,v,x^i)=\exp \Big[\phi + v_0 -i \Big\{p^+v -\sum_{i=1}^8 k_ix^i
+\frac{1}{2p^+} \int du\Big(\sum_{i=1}^8 e^{2v_0-2v_i}k_i^2+m^2e^{2v_0}\Big)
\Big\} \Big],
\label{T2}
\ena
in the background~\p{10D}.
Writing this in our new coordinate system~\p{tf2}, we then find that
due to the identification~\p{ident}, its momentum is quantized:
\bea
p^+ = \e k_1 - \frac{2\pi n}{R}.
\ena
We consider only $n=0$ case, and set
\bea
p^+=\e k_1.
\ena
Requiring its mass squared is positive, we find that
\bea
|k_1| \leq \frac{2u}{\int du e^{2v_0(u)-2v_1(u)}} \e E^-,
\ena
where $E^-$ is the light-cone energy. Similarly
\bea
|k_i| \sim \frac{u}{\Big|\int du e^{2v_0(u)-2v_1(u)}
\int du e^{2v_0(u)-2v_i(u)}\Big|^{\frac12}} \e E^-.
\ena
Examining other components, we find that the momenta in various directions
are of order
\bea
X^+ &:& \e E^-, \nn
X^- &:& \e E^- \times \frac{u}{\int du e^{2v_0(u)-2v_1(u)}}, \nn
X^i &:& \e E^-\times
\frac{u}{\Big|\int du e^{2v_0(u)-2v_1(u)}\int du
e^{2v_0(u)-2v_i(u)}\Big|^{\frac12}}.
\label{od1}
\ena

{}From the metric~\p{st2}, the effective time-dependent string length
$l_s^{\rm eff}$ in the 10-dimensional spacetime is read as
\bea
l_s^{\rm eff}= l_se^{\frac{\phi(\e X^+)}{2}-\frac{v_1(\e X^+)}{2}}
\sqrt{\frac{2\pi l_s}{\e R}}.
\ena
The condition for massive open string to decouple is given by
\bea
E_ol_s^{\rm eff}\ll 1,
\label{op}
\ena
where $E_o$ is typical energy appearing in this theory, which are the
energy scales given in \p{od1}.
This should be satisfied in the $ \e \rightarrow 0$ limit.
We will check that in all examples given below, these can be satisfied.

The condition for gravity to decouple is as follows:
The ten-dimensional effective Newton constant $G_N^{\rm eff}$ is
\bea
G_N^{\rm eff} \sim \frac{4\pi^2l_s^{10}}{\e^2R^2}e^{2\phi (\e X^+)-2v_1(\e X^+)},
\ena
and the strength of the gravity is
\bea
G_N^{\rm eff}E_o^8 \sim \frac{4\pi^2l_s^{10}}{\e^2R^2}
e^{2\phi (\e X^+)-2v_1(\e X^+)}E_o^8,
\label{10y2}
\ena
which should vanish in the limit $ \e \rightarrow 0$.
We will show that this is also valid in all the examples below.
As long as the conditions~\p{op} and \p{10y2} are satisfied,
we may have well-defined matrix string description.

\section{Examples}

Let us now discuss some simple but interesting cases.

\subsection{Linear solutions}

As a simple example, we consider
\bea
v_i(u)=a_i u,~~ (i=0, 1, \cdots, 8), \quad
\phi(u)=a_{9}u.
\ena
We choose $a_9<0$ to have singularity at $u \to -\infty$ and the theory is
strong coupling there. This can be always done by the choice of $u$.
This includes the linear dilaton considered in Ref.~\cite{bb1}. From~\p{cs},
our classical solutions are
\bea
\e X^+ =\left\{ \begin{array}{lcc}
\frac{1}{2a_0}\log{(\frac{\sqrt2 l_s}{R}a_0\tau)}, & & (a_0 \neq 0) \\
\frac{l_s}{\sqrt2 R}\tau, & & (a_0 = 0)
\end{array} \right. ,
\label{ex1}
\ena
and we have
\bea
U_i(X^+)= a_i \e X^+,\quad \phi(\e X^+)= a_{9}\e X^+.
\ena
It is then easy to check that all the conditions~\p{op} and \p{10y2}
are satisfied. The only remaining constraint is the
Einstein equation~\p{bpse}, which simply determines $E'$.

\subsection{Logarithmic solutions}

As a slight generalization of the simple linear background,
we can consider
\bea
v_i(u) = a_i \log |u|, (i=0, 1 ,\cdots,8),~~
\phi(u)=a_{9} \log |u|.
\label{f2}
\ena
We choose $a_9<0$, and the string coupling becomes large near $u \sim 0$.
Here we have more nontrivial background $E'$, and this case includes
spacetime similar to that in Ref.~\cite{time1}. Eq.~\p{cs} gives
\bea
\e X^+ = \Big[\frac{l_s}{\sqrt2 R}(2a_0+1) \tau\Big]^{\frac{1}{2a_0+1}},
\ena
and
\bea
U_i(X^+)=a_i\log{(\e X^+)},\quad
\phi(\e X^X)=a_{9}\log{(\e X^+)},
\ena
both of which are well-defined in the $\e \to 0$ limit.
It can be easily checked that the conditions~\p{op} and \p{10y2} give
\bea
1+a_{9}-a_1 &>& 0, \nn
1 + a_{9} + a_1 + 2 a_i - 4a_0 &>& 0, \nn
3 + a_{9} + 3a_1 + 4a_i - 8a_0 &>& 0,
\ena
for all $i=1,\cdots,8$.
When these are satisfied, we can get the matrix string description.

\subsection{Linear and logarithmic solutions}

As a more nontrivial example, let us consider
\bea
v_i(u) = a_i u,~~(i=0,\cdots,9-d),~~
v_j(u) = a_j \log u,~~(j=10-d ,\cdots,9),~~
\phi(u) = a_9 u.
\label{f2}
\ena
In this case, eq.~\p{cs} gives again \p{ex1} and
\bea
U_i(X^+) = a_i \e X^+ ,\quad
U_j(X^+) = a_j \log{(\e X^+)},\quad
\phi(f) = a_{9} \e X^+,
\ena
are all well-defined in the $\e \to 0$ limit.

The conditions~\p{op} and \p{10y2} give
\bea
a_j > -1.
\ena

\section{Summary and discussion}

It is a very interesting and important problem to examine if and how
string theory can resolve the singularities of our spacetime.
In this paper we have discussed one possible direction in the context
of matrix string theory. Starting from the general class of time-dependent
solutions with $\frac12$ supersymmetry found in \cite{IKO}, we have
formulated how the solutions can be mapped to weak coupling matrix string
theory, where we can examine the behavior of the theory using perturbative
picture.

It appears that we can always map the theory for the $\frac12$ BPS solutions
to matrix string description in the presence of backgrounds, and there is
certain range of validity in this approach. We have also given several examples
in our solutions. Although we have given several examples and there are range
of validity on the parameters in the solutions, we can expect that
there is always valid range for the matrix string description.

Even though there are singularities in the geometrical string picture,
the mapped matrix string theories do not show any violent behavior.
However geometric picture how our spacetime behaves there is not clear.
Our results are consistent with the recent suggestion that
our spacetime is an emergent one. It would be interesting to further
explore the physical picture as to what happens close to the singularities,
and also study other cases where cosmic singularities are tamed.

\section*{Acknowledgement}

We would like to thank H. Kodama and C.-M. Chen for valuable discussions.
The work of NO was supported in part by the Grant-in-Aid for
Scientific Research Fund of the JSPS No. 16540250.

While this paper is being typed, a related paper appeared~\cite{bb15}.


\end{document}